\documentclass{article}

\usepackage{url}
\usepackage{breakurl}

\usepackage{graphicx} 
\usepackage{listings} 

\title{Synchro: Block-generation Protocol to Synchronously Process Cross-shard Transactions in State Sharding}
\author{Takaki Asanuma,
	\thanks{T. Asanuma (asanuma.takaki@fujitsu.com) is a researcher at Fujitsu Limited, Japan.} 
	Takeshi Miyamae,
	\thanks{T. Miyamae (miyamae.takeshi@fujitsu.com) is a senior researcher at Fujitsu Limited, Japan.}
	Yuji Yamaoka,
	\thanks{Y. Yamaoka (yamaoka.yuji@fujitsu.com) is a senior project director at Fujitsu Limited, Japan.}}

\begin{document}

\maketitle

\begin{abstract}
Traditional blockchains cannot achieve the same transaction throughput as Web2, so their use cases are limited.
Therefore, state sharding has been proposed to improve transaction throughput by dividing the blockchain network and managing states and transactions in parallel.
However, Nightshade in the NEAR Protocol, a type of state sharding, provides a rollback protocol to cancel the generation of blocks containing inconsistent transaction results because processing cross-shard transactions (CSTXs) in a 2-phase commit may cause state inconsistency.
We present a new attack that interferes with the generation of new blocks by repeatedly executing CSTXs that certainly causes state inconsistency, causing continuous rollback.
We also propose a block-generation protocol called Synchro to incorporate all the state changes of each CSTX into the same block by coordinating the block prior to approving transactions in each shard.
Synchro eliminates the occurrence of the state inconsistency caused by the CSTXs and the necessity of the rollback protocol.
We use zero-knowledge proof to make Synchro scalable in the global validation phase.
Although the actual overhead of the zero-knowledge proof has not yet been evaluated, we show that Synchro could achieve the same transaction throughput as Nightshade theoretically, depending on the future innovations in zero-knowledge proof techniques.

\end{abstract}

\section{Introduction}
\subsection{Background}
Issues have emerged with traditional Web2 platforms, such as the concentration of personal information and profits in the hands of a few social netwworking service (SNS) companies.
Therefore, Web3 technology, based on blockchain and characterized by its decentralized nature, is attracting increasing attention. 
However, traditional blockchains cannot achieve the same transaction throughput as Web2 SNS.
For example, Twitter has historically achieved a transaction throughput of 143,199 transactions per second (TPS), while Ethereum, the leading smart contract platform, has only achieved about 13 TPS (as of 07/13/2023).\cite{twitter,eth_tps}

Many scalability solutions have been proposed to increase blockchain-transaction throughput.
Layer-1 scalability solutions can be broadly classified into two categories.

\begin{description}
	\item{Scale-up Methods}\\ 
	Solana and Algorand increase transaction throughput by requiring higher hardware requirements, such as for graphics processing units, for validators.\cite{solana,algorand}
	However, the improvement in transaction throughput is also limited by the limited improvement in hardware computing power.
	
	\item{Scale-out Methods}\\
	These methods improve transaction throughput by approving transactions in parallel.
	For example, a directed acyclic graph (DAG), used in IOTA and Avalanche, allows multiple transactions to be approved in parallel because the blocks are connected in a DAG-like fashion.\cite{popov2018tangle,ava}
	State sharding, which is used in Ethereum 2.0 and the NEAR Protocol, applies sharding technology, a database-partitioning method, to blockchains.\cite{eth2.0,nightshade,wang2019sok,han2021security,mizrahi2020blockchain,hashim2022sharding}
	State sharding improves transaction throughput by dividing blockchain states and validators into subsets called shards and approving transactions in parallel on several shards.
\end{description}

The transaction throughput of scale-up methods depends on the hardware performance of the validator.
Scale-out methods of DAG parallelize only the transaction-approval process and does not partition the state management.
State sharding divides the blockchain state, allowing for parallelizing a theoretically almost unlimited number of transaction approvals.\footnote{There are variants of sharding blockchains that do not split and manage up to state such as Zilliqa and Elastico.\cite{zilliqa,elastico}}
Therefore, this paper focuses on scale-out methods in terms of their potential concerning transaction throughput.

In state sharding, state inconsistency can occur as a result of failed state changes.
State sharding involves transactions processed across several shards, called cross-shard transactions (CSTXs).\cite{hashim2022sharding,wang2019sok,han2021security}
State sharding, such as Ethereum 2.0 and NEAR Protocol, uses a 2-phase commit (2PC) to process CSTXs.\cite{han2021security,nightshade}
However, 2PC changes multiple states serially and asynchronously, which can compromise state consistency.

There are two solutions to this problem.
The first is to send an additional transaction to resolve state inconsistency due to CSTX inconsistency.\cite{eth_addtx}
However, if the processing of the additional transaction fails, further additional transactions are required, making the protocol more complex.
The second solution is to roll back the ledger to undo the generation of blocks containing transactions that do not meet consistency.\cite{nightshade}
This solution avoids the protocol complications associated with further additional transactions.
However, an attack can sabotage block generation by exploiting rollback.
A denial of service (DoS) attack that repeatedly executes CSTX, which generates state inconsistencies, could prevent new blocks from being generated if successive rollbacks occur.

\subsection{Contributions}
Our contributions are summarized below.

\begin{itemize}
	\item We present a new attack that interferes with the generation of new blocks by repeatedly executing CSTXs that certainly causes state inconsistency, causing continuous rollback.
	
	\item We propose a block-generation protocol called Synchro to incorporate all the state changes of each CSTX into the same block by coordinating the block prior to approving transactions in each shard.
	We also show that this protocol could achieve the same transaction throughput as NEAR Protocol theoretically.
\end{itemize}

\subsection{Organization of This Paper}
Chapter 2 describes prior work on state sharding.
Chapter 3 introduces a new attack that interferes with the generation of new blocks by repeatedly executing CSTXs that certainly causes state inconsistency, causing continuous rollback.
Chapter 4 presents a block-generation protocol called Synchro to incorporate all the state changes of each CSTX into the same block by coordinating the block prior to approving transactions in each shard.
Chapter 5 summarizes our evaluation of the security and scalability of Synchro.
Finally, Chapter 6 presents the conclusions of this work.

\section{Previous Work}
\subsection{Cross-Shard Transaction in State Sharding}
State sharding defines a CSTX as a transaction that changes states in several shards.\cite{hashim2022sharding,wang2019sok,han2021security}
Since a validator is only expected to manage the states in the shard to which it belongs, there is no validator that manages all the states.
In other words, one validator cannot handle any CSTXs by itself.

In Ethereum 2.0, 2PC is used to handle CSTX.\cite{han2021security,nightshade}
In 2PC, the state is changed serially on the two shards.
The validator on a shard approves the former half of state changes of a CSTX sends the execution result to the validator of the next shard.
The transaction information required to approve the latter half of state changes, including the CSTX and the former execution result,  is called a receipt. 

\subsection{Challenges in Cross-Shard Transaction}
In 2PC of Ethereum 2.0, an inconsistency in the blockchain state occurs if the former state changes were approved but the latter ones were not.
In such a case, a new independent transaction is automatically invoked to resolve the inconsistency.\cite{eth_addtx}
However, there is a possibility that the additional transaction may also fail, exponentially increasing the number of rare anomaly cases that must be assumed in the system design.

The NEAR Protocol's Nightshade also uses 2PC, but the white paper proposes a rollback protocol for resolving inconsistencies in the blockchain state.\cite{nightshade}
The rollback protocol eliminates the need to invoke additional transactions, making the NEAR Protocol less complicated.

\section{Block Generation Interference Attack}
In NEAR Protocol, a DoS attack that repeatedly executes CSTXs, which causes state inconsistency, could result in a series of rollbacks, which could interfere with block generation.
This chapter introduces an attack that exploits the rollback implemented in the NEAR Protocol to halt block generation semi-permanently.
We first provide an overview of the NEAR Protocol then summarize the attack and its steps.

\subsection{Overview of Nightshade}
Before describing the block-generation interference caused by DoS attacks in Nightshade, this section describes the block-generation protocol in Nightshade.
Figure ref{fig:near} shows an overview of the block-generation flow of the NEAR Protocol.
The user sends a transaction to a validator in the shard to which he or she belongs.
Several validators called producers belong to each shard, and one producer validates the transaction received from the user in turn.
When processing CSTX, the producer also creates a receipt.
The producer then compiles the approved transactions and receipts into sub-blocks called chunks and distributes the created chunks to other validators inside and outside the shard.

In Nightshade, smart contracts, like states, belong to one of the shards.\cite{contract}
When one account calls a function of a contract in another shard, the call is made via a receipt, as in 2PC.\cite{cross-call}

Nightshade takes into account each chunk as one block, but the validator holds only the chunks of the shard to which it belongs and does not centrally manage the entire block.
A CSTX is processed using 2PCs, so receipts will be stored in the block after the block from which the CSTX was taken.

A rollback is also triggered by a challenge by a validator that discovers an incorrect transaction, such as one that causes inconsistency in the state, and all validators cancel the target transaction from their ledgers.\footnote{NEAR Protocol rollback is described in the white paper.\cite{nightshade} But as of June 2023, the details of the rollback are unknown as it has not yet been implemented.}

\begin{figure}[t]
	\begin{center}
		\includegraphics[width=\linewidth]{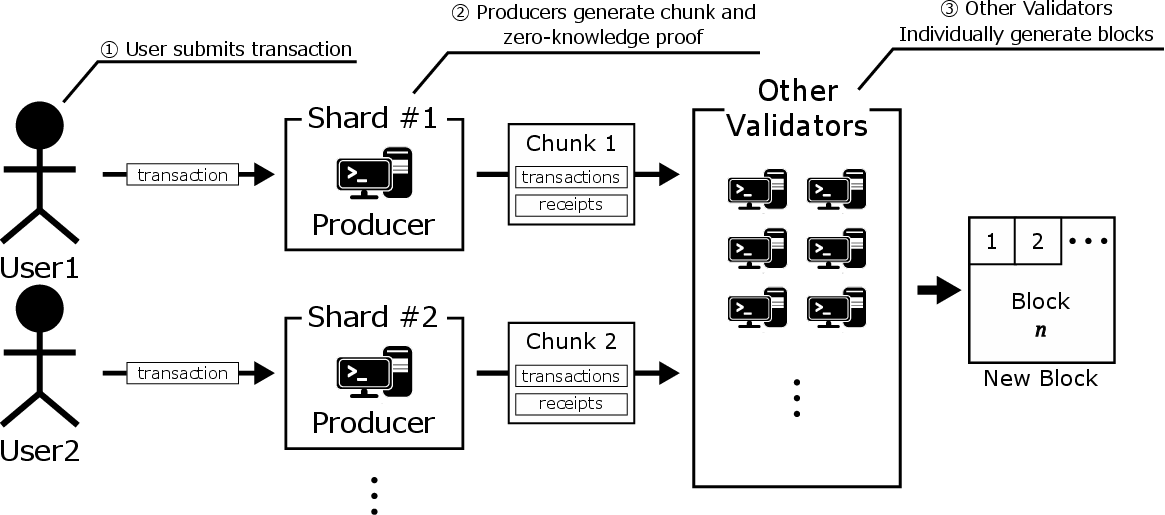}
		\caption{Overview of NEAR Protocol's block-generation flow}
		\label{fig:near}
	\end{center}
\end{figure}

\subsection{Attack Summary}
The attacker uses a contract with a denial function that rejects calls by receipts under certain conditions.
The attacker deploys a homemade attack contract and sends a CSTX that causes the invocation to be rejected by the contract, thus rejecting the receipt and causing inconsistency in the state.
It is also possible to cause a series of rollbacks by repeatedly sending CSTX that cause state inconsistency.
All users can carry out this DoS attack, and by using this attack, it is possible to interfere with block-generation at any given time.

\subsection{Attack Procedure}
The attack procedure is shown below.

\begin{description}
	\item{Step 1: Develop smart contracts to be used in the attack}\\
	The attacker develops a smart contract for the attack that implements the following functions.
	An example implementation of the contract is shown in Appendix\ref{contract}.
	\begin{itemize}
		\item Functionality that enables tokens to be sent to this contract
		
		\item If this contract holds more than 1NEAR\footnote{NEAR represents the monetary unit of the native tokens issued on the NEAR Protocol network.}, a function is executed to reject the call using the revert function and cancel the transaction process in the middle.
	\end{itemize}
	
	\item{Step 2: Deploy the contract}\\
	The attacker deploys the developed contract.

	\item{Step 3: Send 1NEAR to the contract}\\
	The attacker executes the deposit function using an account belonging to a different shard from the attacking contract and sends 1NEAR.
\end{description}

Step 3 causes the attacker's contract to hold more than 1NEAR, so the revert function rejects the receipt call and cancels the state change resulting from the contract's deposit.
Therefore, only the CSTX will be approved, resulting in an inconsistency in the blockchain state, and a rollback will occur as a response to the abnormal system.
In this attack, the 1NEAR sent by the rollback will be refunded, so it is possible to cause a rollback with only the cost of gas.
Repeating this attack makes it possible to interfere with new-block generation by ordinary users. \footnote{As of June 2023, the validity of the attack has not been confirmed through demonstration testing, as it has not yet been implemented}.

\section{Proposed Consistent Block-generation Protocol}
\subsection{Construction of Synchro}
We propose a block-generation protocol that incorporates all the state changes of each CSTX into the same block by coordinating the block prior to approving transactions in each shard.
We explain the differences from the NEAR Protocol.

\begin{description}
	\item{Introduction of global validator}\\
	To prevent inconsistency in the blockchain state, a role is needed to ensure that CSTXs and receipt approvals are completed in a same block.
	Therefore, Synchro adds a new role called global validator.
	The global validator does not belong to any shard and does not maintain a state.
	The global validator receives chunks from the producers of each shard and validates all chunks using zero-knowledge proofs to ensure that CSTX and receipt approvals are complete in a same block.
	Thus, the global validator can prevent inconsistency in the blockchain state by verifying that the CSTX and receipt authorization is complete within a same block.
	
	\item{Creation of zero-knowledge proofs by producers}\\
	The global validator validates a chunk of all shards to ensure the producer validates correctly.
	Therefore, the global validator must maintain the state of all shards, which requires storage costs and computational load associated with maintaining and updating the state.
	Therefore, to achieve state sharding, Synchro produces a zero-knowledge proof for the producer to prove the validity of the chunks.\cite{miyamae2023zgridbc}
	Zendoo, proposed in a previous work, is an extension framework for the blockchain developed by Horizon.\cite{garoffolo2020zendoo}
	It enable interoperability between the main chain and side chains.
	When a side chain sends a transaction to the main chain, it uses zero-knowledge proofs to prove the validity of the transaction, allowing the main chain to validate the transaction without determining the state of the side chain.
	In Synchro, to allow the global validator to validate chunks statelessly, the producer creates a zero-knowledge proof, similar to Zendoo, to prove the validity of the chunk's state changes.
	Thus, the global validator's storage cost and computational load are reduced.
	
	\item{Introduction of coordinator}\\
	If several producers in each shard were free to create chunks, the global validator would need to combine chunks so that CSTXs and receipts could be processed within a same block. 
	However, a considerable number of chunk combinations would have to be validated.
	Therefore, Synchro adds a new role, called coordinator, to limit the number of chunk combinations.
	The coordinator does not belong to any shard and does not maintain a state.
	The coordinator receives transactions from users and generates blocks.
	If a CSTX is included in the chunk that makes up the block, the producer creates the corresponding receipt conventionally.
	However, in Synchro, the coordinator creates the corresponding receipt and stores it in a chunk of the same block.
	Thus, the coordinator generates the entire block, and the global validator only needs to verify the combination of chunks created by the same coordinator.
\end{description}

\subsection{Block-generation Flow of Synchro}
\begin{figure}[t]
	\begin{center}
		\includegraphics[width=\linewidth]{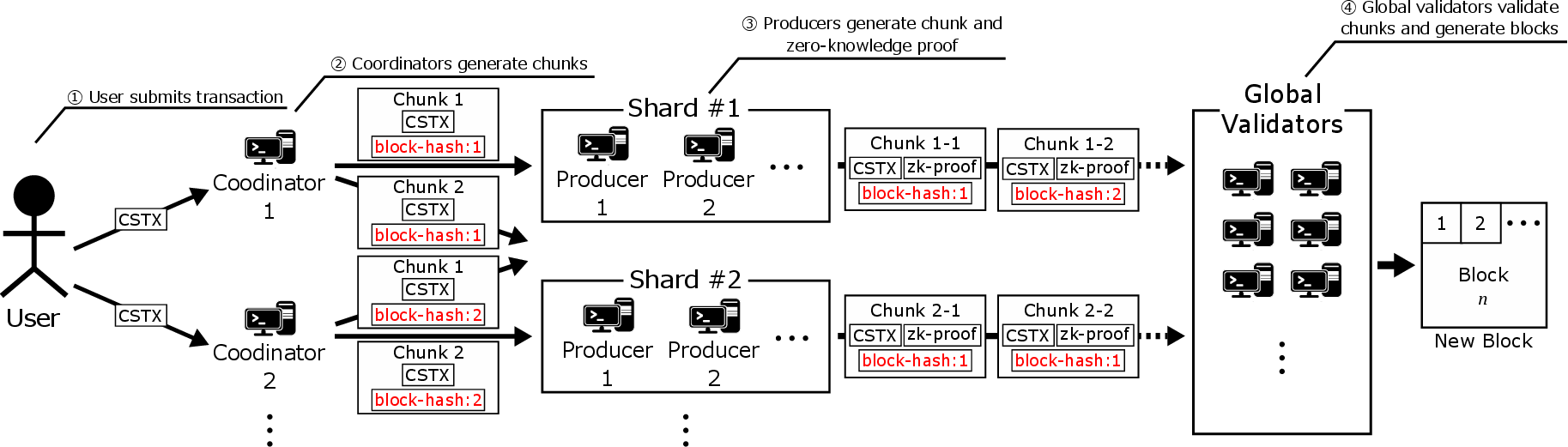}
		\caption{Overview of Synchro's block-generation flow}
		\label{fig:scheme}
	\end{center}
\end{figure}

Figure \ref{fig:scheme} shows an overview of the Synchro's block-generation flow.
This section describes the procedure of Synchro in chronological order.

\begin{description}
	\item{1. Transaction submission by user}\\
	The user sends the created transaction to all coordinators.
	
	\item{2. Block-generation by coordinator}\\
	Coordinators receive transactions from users and generate blocks independently.
	However, as multiple coordinators generate blocks independently, the global validator receives many chunks and requires help in identifying the correct combination.
	Therefore, after generating a block, the coordinator calculates a block hash and stores it in the chunk.
	In this way, the global validator can rely on the block hash to combine chunks and reconstruct the block.
	
	\item{3. Chunk verification and zero-knowledge proof production by producer}\\
	The producer selects one coordinator and requests chunks and block hashes.
	The received chunks are verified using a state to confirm the validity of the chunks.
	If the chunk is approved, the producer creates a zero-knowledge proof of the chunk's validity to allow the global validator to validate the chunk statelessly.
	The producer then receives a request from the global validator and sends the chunk and zero-knowledge proof.
	
	\item{4. Chunk validation and block-generation by the global validator}\\
	The global validator designates a producer for each shard to request chunks and zero-knowledge proofs.
	It combines the chunks from the received chunks by relying on the block hash and reconstructing the block.
	By reconstructing the block generated by the coordinator, it is possible to easily search for a combination of chunks that allows CSTX and receipt approval to be completed within a same block.
	The global validator also verifies the validity of state changes using zero-knowledge proofs for all chunks that comprise the reconstructed block.
	If all chunks are approved, the global validator combines the chunks to generate a block.
\end{description}

Generating blocks by the global validator may result in multiple types of blocks.
Therefore, the blockchain may fork, but Synchro will determine the legitimate chain in accordance with the fork-choice rule.

\subsection{Incentives for each role}
Synchro prevents fraud by multiplexing each role.
However, such fraud cannot be prevented unless a sufficient number of honest roles are gathered.
Therefore, we introduce incentives to induce each role to participate actively.

Specifically, each role will be given a digital signature for the data it generates so that it can later be verified by an unspecified number of participants who generated the data.
An incentive rule is also added in which all participants involved in the generation of the approved block will be granted a token.
Thus, each role is incentivized to follow the protocol for incentive acquisition.
Since incentive acquisition is a joint effort of all roles, producers will come to trust the coordinator that has proposed several blocks that have been approved in the past and will receive chunks from that coordinator.
The global validator will also come to trust the producer that has proposed several chunks for previously approved blocks and will preferentially validate chunk combinations from that producer.
Thus, malicious coordinators and producers that attempt to propose incorrect blocks or chunks will eventually be weeded out, as honest producers and global validators will ignore them.

\subsection{Effectiveness of Synchro}
In conventional state sharding, CSTX is handled using 2PC.
However, 2PCs process CSTXs and receipts serially, resulting in the inconsistency of blockchain states.
Therefore, Synchro introduces a new coordinator and global validator.
Synchro can achieve 1PC without sacrificing efficiency and decentralization by having the coordinator propose candidate blocks so that CSTX and receipt approvals can be completed in a same block and having the global validator verify the consistency of CSTXs and receipts. 
Therefore, Synchro does not require rollback for inconsistencies in the blockchain state, making DoS attacks that cause continuous rollback impossible.

\section{Evaluation}
\subsection{Security}
While Synchro adds new roles, we ensure that cheating by each role does not occur.

\begin{description}
	\item{Coordinator}\\
	It is possible that a malicious coordinator could intentionally ignore transactions involving a particular user.
	If this disregard were carried out, the transactions of a particular user would be interfered with.
	However, since Synchro multiplexes coordinators, other coordinators store that transaction in a block even if an attacker ignores a particular transaction.
	Thus, it would be impossible for an attacker to interfere with transactions involving a particular user.
	
	A malicious coordinator could generate a block containing only CSTX or receipts.
	If that block were approved, only one of the states would be changed, creating the possibility of lost tokens and double payments.
	However, in Synchro, the global validator must verify that CSTX and receipt approval are completed in a same block.
	Therefore, an attacker cannot change only one of the states in the CSTX in a same block, and no token loss or double payment can occur due to inconsistency in the blockchain states.
	
	\item{Producer}\\
	A malicious producer could create a malformed chunk and send it to the global validator instead of the chunk created by the coordinator.
	A DoS attack by sending malformed chunks would enable an attacker to interfere with the generation of new blocks by ordinary users.
	However, since Synchro enables each producer to select and propose chunks to the global validator independently, the global validator can successfully generate blocks if at least one producer belonging to the same shard as the attacker operates in accordance with the protocol.
	Thus, as long as there is an honest producer in each shard, it will be impossible to interfere with block generation.
	
	\item{Global Validator}\\
	A malicious global validator could generate blocks of incorrect chunk combinations.
	For example, token loss or double payments could occur if an attacker generates a block that does not contain CSTX and receipts simultaneously.
	However, the Synchro is supposed to determine the authorized chain in accordance with the fork-choice rule.
	Thus, if most global validators are honest, fraudulent blockchains will be eliminated.
\end{description}

\subsection{Scalability}
The new roles and zero-knowledge proofs in Synchro may reduce transaction throughput compared with Nightshade.
Therefore, we theoretically compared Nightshade and Synchro in terms of transaction throughput.

\begin{description}	
	\item{Changes associated with the introduction of coordinators}\\
	In Synchro, a new coordinator is introduced, which causes a delay in transaction processing due to the overhead associated with it.
	However, the processing of the coordinator can be pipelined with the processing of other roles.
	Therefore, if the coordinator can generate a block within the block time, the overhead associated with the introduction of the coordinator will not affect transaction throughput.
	Since the coordinator's processing is only central-processing-unit computation and network communication, it should be possible to complete the processing during the block time.
	
	\item{Changes associated with the introduction of zero-knowledge proofs}\\
	Since Synchro introduces a new zero-knowledge proof, there is an associated overhead.
	However, when the average block time of Nightshade is $t_{block}$, chunk-verification time by the producer is $t_{chunk}$ and zero-knowledge-proof production time is $t_{zk_p}$, Formula (\ref{eq:zk}) is satisfied.
	The overhead does not affect scalability.
	
	\begin{equation}
		t_{block} \geq t_{chunk} + t_{zk_p} \label{eq:zk}
	\end{equation}
	
	As of July 7, 2023, $T_{block}$ is about 1s.\cite{blocktime}
	Prior research has found that the time to create a zero-knowledge proof using the simplest generation circuit is $t_{zk_p}=0.41$s\cite{miyamae2023zgridbc}.
	Therefore, if Formula (\ref{eq:zk}) is satisfied, the block time of Synchro will not be reduced compared with Nightshade.
	
	Assuming that 100,000 TPS is achieved with 1000 shards, 100 transactions per chunk would be required (the upper limit in the specification is 1000 transactions per chunk).
	Assuming that a zero-knowledge proof is generated for each transaction, a simple calculation shows that it would take 0.41s$\times$100 = 41s, a performance degradation compared with Nightshade.
	Therefore, it is necessary to keep the proof-creation time within 1s by adopting a policy of creating one zero-knowledge proof per chunk, designing the proposition circuit as simple as possible, and adopting the most efficient zero-knowledge-proof protocol.
	
	\item{Changes associated with the introduction of global validators}\\
	In Nightshade, only the producer generates the block.
	However, Synchro adds verification of the block by a global validator, which causes a delay in transaction processing due to the associated overhead.
	However, the processing of the producer and global validator can be pipelined.
	If $t_{chunk}$ denotes the chunk-validation time by the producer, $t_{zk_p}$ the zero-knowledge-proof production time, $t_{zk_v}$ the chunk validation by the global validator, and $s$ the number of shards, then Formula (\ref{eq:global}) is satisfied The overhead of adding the global validator does not affect transaction throughput.
	
	\begin{equation}
		t_{chunk} + t_{zk_p} \geq s \times t_{zk_v} \label{eq:global}
	\end{equation}
	
	In a previous work, the verification time with zero-knowledge proofs using the simplest generation circuit was $t_{zk_v}=0.0043$s.\cite{miyamae2023zgridbc}
	For $s=100$, $s \times t_{zk_v}=0.43$s.
	Therefore, given that $t_{zk_p}=0.41$s, it is possible to satisfy Formula (\ref{eq:global}) depending on $t_{chunk}$.
\end{description}

\section{Conclusion}
We presented a new attack that interferes with the generation of new blocks by repeatedly executing CSTXs that certainly causes state inconsistency, causing continuous rollback.
We proposed a block-generation protocol called Synchro to incorporate all the state changes of each CSTX into the same block by coordinating the block prior to approving transactions in each shard.
In this protocol, we showed that avoiding DoS attacks that repeatedly execute CSTXs is possible, which causes state inconsistency.
We also showed that this protocol could theoretically achieve the same transaction throughput as NEAR Protocol, depending on technical and future innovations in zero-knowledge proofs.


\section*{Author Contributions}
All authors contributed equally to this work.

\section*{Conflict of Interest}
All authors have affirmed they have no conflicts of interest as described in Ledger’s Conflict of Interest Policy.



\bibliographystyle{ledgerbib}
\bibliography{./reference}



\appendix
\setcounter{section}{0}
\lstset{
	basicstyle={\ttfamily},
	identifierstyle={\small},
	commentstyle={\smallitshape},
	keywordstyle={\small\bfseries},
	ndkeywordstyle={\small},
	stringstyle={\small\ttfamily},
	frame={tb},
	breaklines=true,
	columns=[l]{fullflexible},
	numbers=left,
	xrightmargin=0pt,
	xleftmargin=5pt,
	numberstyle={\scriptsize},
	stepnumber=1,
	numbersep=1pt,
	lineskip=-0.5ex
}

\section{Attack Contract Implementation}
\begin{lstlisting}[caption=Example of Attack Contract Source Code(RUST), label=contract]
	use near_sdk::borsh::{self, BorshDeserialize, BorshSerialize};
	use near_sdk::{env, near_bindgen, Balance};
	
	#[near_bindgen]
	#[derive(Default, BorshDeserialize, BorshSerialize)]
	pub struct Contract {
	}
	
	#[near_bindgen]
	impl Contract {
		#[payable]
		pub fn send() {
			assert!(env::account_balance() > 1e24, "Over");
		}
	}
\end{lstlisting}


\end{document}